%% file: _main.tex
\documentclass[manuscript,nonacm]{acmart}

\AtBeginDocument{%
  \providecommand\BibTeX{{%
    \normalfont B\kern-0.5em{\scshape i\kern-0.25em b}\kern-0.8em\TeX}}}

\setcopyright{acmcopyright}
\copyrightyear{2018}
\acmYear{2018}
\acmDOI{XXXXXXX.XXXXXXX}

\acmConference[Conference acronym 'XX]{Make sure to enter the correct
  conference title from your rights confirmation emai}{June 03--05,
  2018}{Woodstock, NY}
\acmBooktitle{Woodstock '18: ACM Symposium on Neural Gaze Detection,
 June 03--05, 2018, Woodstock, NY} 
\acmPrice{15.00}
\acmISBN{978-1-4503-XXXX-X/18/06}

\usepackage{xcolor}
\usepackage{tcolorbox}
\usepackage{multirow}
\usepackage{xr}

\newcommand{\edit}[1]{\textcolor{black}{#1}}
\newcommand{\finaledit}[1]{\textcolor{black}{#1}}

\usepackage{titlesec}
\usepackage{enumitem}

\newcommand{\recsection}[3]{
    \paragraph{\textit{#1}} {#2}
    \ifthenelse{\equal{#3}{}}{}{\newline\hspace*{2em}\textit{Consider}: #3}
}

\begin{document}

\title[AURA: Responsible AI Content Work]{AURA: Amplifying Understanding, Resilience, and Awareness for Responsible AI Content Work}

\author{Alice Qian Zhang}
\email{zhan6698@umn.edu}
\affiliation{%
  \institution{University of Minnesota}
  \country{USA}
}

\author{Judith Amores}
\email{judithamores@microsoft.com}
\author{Mary L. Gray}
\email{mlg@microsoft.com}
\author{Mary Czerwinski}
\email{marycz@microsoft.com}
\author{Jina Suh}
\email{jinsuh@microsoft.com}
\affiliation{%
  \institution{Microsoft Research}
  \country{USA}}

\renewcommand{\shortauthors}{Zhang, et al.}

\include{sections/0_abstract}





\maketitle

\input{sections/1_intro}
\input{sections/2_background}
\input{sections/3_phase1_methods}
\input{sections/4_phase1_findings}
\input{sections/6_phase2}

\input{sections/7_overall_discussion}

\input{sections/8_conclusion}

\bibliographystyle{ACM-Reference-Format}
\bibliography{_references}
\end{document}

%% file: sections/0_abstract.tex
\begin{abstract}
\edit{Behind the scenes of maintaining the safety of technology products from harmful and illegal digital content lies unrecognized human labor. }
The recent rise in the use of generative AI technologies and the \finaledit{accelerating demands to meet} responsible AI (RAI) aims necessitates \finaledit{an increased focus on the labor behind such efforts in the age of AI}. This study investigates the nature and challenges of content work that supports RAI efforts\finaledit{, or ``RAI content work,'' that} span content moderation, data labeling, and red teaming -- through the lived experiences of content workers. We \finaledit{conduct a }formative survey and \finaledit{semi-structured }interview studies to develop \finaledit{a conceptualization of RAI content work and a subsequent }framework of recommendations for providing holistic support for content workers. We validate our recommendations through a series of workshops with content workers and derive considerations for and examples of implementing such recommendations. We discuss how our framework may guide future innovation to support the well-being and professional development of the RAI content workforce.
\end{abstract}

%% file: sections/1_intro.tex
\section{Introduction}

On July 21, 2023, the United States White House released a statement detailing the voluntary commitments of companies leading in developing artificial intelligence (AI)~\cite{The_White_House_2023}.
These commitments include promises to ensure AI systems are safe through ``internal and external testing'' before their introduction to the public. 
Such promises subsequently raise concerns about how human expertise is being recruited and supported in this type of testing. Thus in this paper, we explore how to best support people engaging in work practices that ensure ethical and safe AI products.

\finaledit{We define those practices as \textbf{Responsible AI (RAI) content work}, 
which involves generating, reviewing, or reasoning about digital content with the goal of ensuring safety and ethical standards in AI systems ~\cite{OpenAI_2023}. In this paper, we focus on three key aspects of RAI content work to scope our study: content moderation, data labeling, and the emerging practice of red teaming. 
These areas are critical to ensuring the ethical and responsible development of contemporary AI systems. It is important to note, however, that individual RAI content workers may engage in a multitude of these activities, reflecting the multifaceted nature of their role in supporting responsible AI development.} 
\finaledit{Regardless of the specific activities workers engage with,} the support for human efforts behind these initiatives is often overlooked despite the importance of the work conducted \cite{wohn2019volunteer, dosono2019moderation, pinchevski2023social, schopke-gonzalez_why_2022, steiger_psychological_2021, gray2019ghost}.
Without a comprehensive understanding \finaledit{of these efforts, we may see history repeat itself with content work facing challenges of invisibility of the workforce and a lack of well-being support crucial to workers. }

Prior human-computer interaction (HCI) literature on harmful content exposure within content moderation has surfaced key challenges of developing psychological symptoms such as anxiety, depression, and burnout within populations ~\cite{wohn2019volunteer, dosono2019moderation, pinchevski2023social, schopke-gonzalez_why_2022, steiger_psychological_2021}.
However, empirical data on how these challenges manifest in other types of content work and factors unique to RAI (e.g., sudden increases in content volume due to interest in AI integration) remains limited.
Studies have also explored using technologies to mitigate harmful content exposure and treat symptoms \cite{holman_medias_2014, das_fast_2020, karunakaran2019testing,steiger2022effects}, but were \finaledit{limited to primarily image and video-based content that does not cover the full spectrum of types of exposure in all types of content work}.
Recent calls within Computer-Supported Cooperative Work (CSCW) advocate examining the transformation of human labor within AI systems \cite{sheehan2023making, cheon2023powerful}. 
In this context, we investigate the emergence of \finaledit{RAI content work} as a new form of digital labor and the potential disruptions generative AI may bring to the digital content ecosystem, raising uncertainty about the impact on those maintaining AI system safety.

Previous studies within CSCW have examined content moderation challenges specific to end-user communities ~\cite{Zhang2023, chancellor2016thyghgapp, rubya2017video} and platforms ~\cite{reddy2023evolution, han2023hate}. 
However, the challenges related to the well-being and work quality of content workers employed and working with AI systems have yet to be explored in depth. 
To address this gap, we ground our study in the lived experiences of self-identifying content workers engaged in various activities with and around content. 
\finaledit{We aim} to highlight the need to evaluate challenges content workers face amidst growing AI-related content demands and to inform \finaledit{future} practices \finaledit{of content work in the age of AI}.

\begin{figure*}
    \centering
    \includegraphics[width=\textwidth]{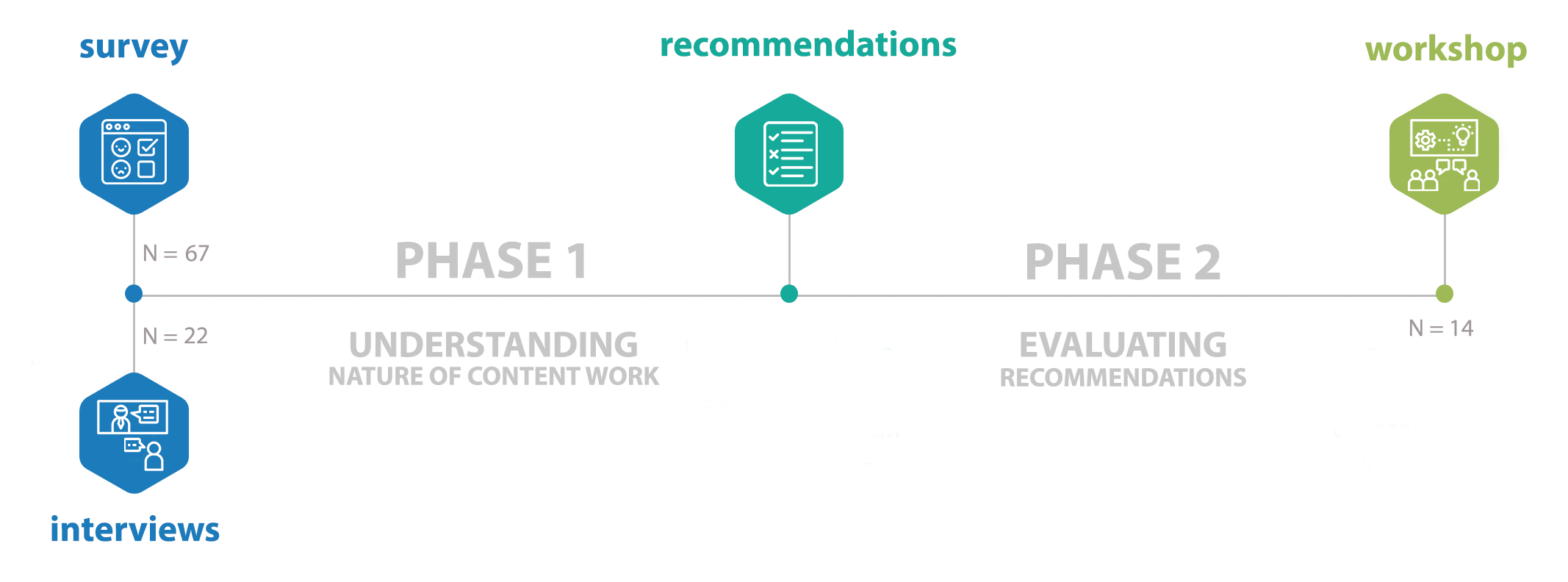}
    \caption{Flow of our two-phase study. In the first phase, we conducted a survey study (N=67) and an interview study (N=22) to understand the nature of content work. From these insights, we developed a set of recommendations to improve content worker well-being. In the second phase, we validated the challenges we discovered and our recommendations to address those challenges, within the AURA framework that organize those recommendations, through interactive workshops (N=14).}
    \label{fig:studyflow}
\end{figure*}

 We take a comprehensive approach, examining \finaledit{all types of content work from content moderation to red teaming} through a two-phase study (see \autoref{fig:studyflow} for the study flow).
In the first phase, we provide empirical insights from surveys and interviews on the nature of content work ~(\ref{methods:rq-nature}) and the challenges content workers face ~(\ref{methods:rq-challenge}). 
We illustrate the multi-faceted nature of content work, detailing findings from the main factors that constitute it: workers' roles, types of content that workers are exposed to, protective tools they use, impacts of engaging with content, and practices for collaboration. 
Building off of these insights, we surface challenges about misconceptions about the realities of content work, shortcomings of tools and metrics, failures of workplace support, and barriers to career growth. 
These challenges informed our proposal of a framework for amplifying understanding, resilience, and awareness (AURA) for RAI Content Workers comprised of four categories: \textit{recruitment, tooling, adaptive wellness, and retention}. 
In the second phase, we further revise our recommendations through validation workshops that surface challenges and considerations for the applications of these recommendations in our framework.   
Overall, our study informs future improvements in the design of content work, developments that can support the well-being of workers, and progress in defining the professional identity and growth of the RAI workforce.

%% file: sections/2_background.tex
\section{Related Work}
In the following section, we delve into relevant existing literature to provide context for our study.
\edit{We first provide a background of how RAI content work is conceptualized, tracing the origins of content-related endeavors} from \edit{annotation and moderation of harmful online content} to the formulation of our definition of content work within RAI\edit{, as many forms of content work are dedicated to shielding users from potentially harmful AI content and behaviors}.
Subsequently, we \edit{summarize} the well-documented challenges that \edit{content workers dealing with harmful content or behaviors may} encounter, setting the stage for exploring the new obstacles and difficulties content workers face \edit{during RAI efforts}.
\edit{Lastly}, we \edit{highlight prior work on targeted support measures for content workers, from tooling and automation to organizational interventions to motivate renewed emphasis on holistic examination of support for all types of content-related work necessitated by the AI boom}. 

\subsection{Conceptualizing Content Work \edit{for Responsible AI}}
\edit{
Within the landscape of RAI, ``content work'' (or ``data enrichment work''~\cite{PAI_2021}) encompasses any form of labor that requires human review, judgment, or intelligence on a digital content or data that may be used to ensure the safety and responsibility during AI or ML model development. 
This includes data labeling, annotations, or validation as well as content moderation, human feedback, or corrections. 
}
\edit{
The labor of annotating, reviewing, or moderating potentially harmful content or behaviors from within technology-facilitated spaces or interactions could be traced back to early online communities of unpaid \textit{volunteers}~\cite{schopke-gonzalez_why_2022} that enforced community norms and rules~\cite{roberts2017content, matias2019civic} across platforms like Discord~\cite{seering2022pride}, Twitch~\cite{wohn2019volunteer}, Reddit~\cite{li2022all}, and Facebook~\cite{gibson2023teams}.
In addition, \textit{end-users} who are not formal community members may also engage in content moderation by reporting violations~\cite{crawford2016flag, jhaver2023personalizing}, adjusting moderation preferences~\cite{jhaver2023users, geiger2016bot, keller2021future}. 
As the popularity of social media platforms grew, this labor evolved into a structured and industrialized form with human workers in various ways.
\textit{Crowdsourced content workers} are those who are recruited to complete small human intelligence tasks, such as labeling content on Amazon MTurk~\cite{morrow2022emerging, das_fast_2020,hettiachchi2019towards}.
\textit{Commercial content workers} (i.e., commercial content moderators or CCMs~\cite{roberts2017content}) are a dedicated workforce employed in some full- or part-time capacity.
}
\edit{
Recently, the demand for commercialized content work seems to be on the rise. 
For example, content moderation, as an already booming market~\cite{Fact_MR_2024}, is now met with an increased demand for moderating AI-generated content~\cite{Avasant_2024}.
}

\edit{
However, the traditional data labeling and content moderation are not the only forms of content work on the rise. 
Originating as a military strategy~\cite{Zenko_Haass_2015, longbine2008red}, red teaming has since been applied to the field of security~\cite{zenko2015red, abbass2011computational, wood2000red, nguyen2022cti4ai, kazim2020ai} and more recently within RAI as a mandate~\cite{The_White_House_2023} and a method~\cite{ganguli_red_2022, Casper2023ExploreEE} to safeguard AI deployments.
Similarly to the traditional forms of content work, RAI red teaming has been seen in the public domain with volunteers~\cite{Ai_Red_Team_2023, ai-red-team-defcon,Chowdhury_2023} and crowdworkers~\cite{ganguli_red_2022}, and seen in commercial domains with full-time operators~\cite{Microsoft_2023, OpenAI_2023, FrontierModelForum}. 
RAI red teaming also shares similar goals as other forms of content work where it broadly seeks to identify potentially harmful capabilities or outputs from AI systems~\cite{brundage_toward_2020, solaiman_evaluating_2023}, with the exception that it often involves a structured and systematic adversarial testing~\cite{Microsoft_2023,FrontierModelForum}.
Despite such recent popularity, there is limited research about best practices for RAI red teaming or the experiences of people who identify as conducting RAI red teaming~\cite{Singh_Blili-Hamelin_Metcalf_2023,Kumar_2023, Fabian_2023,  Microsoft_2023,Friedler_2023}. 
Because of the adversarial nature of RAI red teaming work, it should not be prematurely assumed as equal to or distinct from the other forms of content work.
}

\edit{
Our work proposes a renewed and holistic examination that is inclusive of data labeling, content moderation, and red teaming as an emerging form of content work.
Such examination is important for the design a supportive framework for RAI content work and timely because all such forms of content work may increase in demand and become further commercialized as AI becomes prevalent.
In doing so, we leverage past learning from well-known professions, such as content moderation, and incorporate the lived experiences of all types of content workers to inform recommendations for existing and emerging RAI practices.
}

\subsection{Well-being Challenges for Content Workers \edit{from Harmful Content Exposure}}\label{sec:background-challenges}
\edit{The labor of annotating, moderating, and testing data for RAI} poses specific challenges concerning the well-being of content workers, as they consistently face the risk fo encountering potentially harmful content. 
Prior research \edit{highlighted} the emotional toll on content \edit{workers along} a wide spectrum of \edit{psychological} impacts.
\edit{Impacts include} symptoms of secondary trauma\edit{, burnout}, a sense of undervaluation for their contributions, feelings of privacy infringement~\cite{wohn2019volunteer, dosono2019moderation, pinchevski2023social, schopke-gonzalez_why_2022} \edit{as well as alterations to} their belief systems~\cite{newton_trauma_2019, Stackpole_2022, Douek_2021} \edit{and the development of} post-traumatic stress disorder (PTSD)~\cite{Michel2018ExContentMS, ruckenstein_re-humanizing_2020, Dwoskin_2019, arsht_2018_human}.
\edit{Unfortunately, much of the prior research on content workers remain in the volunteer or crowdwork domain.}
\edit{On the commercial domain}, there is limited knowledge about content workers' challenges and the types of support they typically receive in organizations~\cite{roberts2017content, carmi2019hidden}. 
For example, CCM research often relies on end-user perceptions due to limited transparency from platforms~\cite{suzor2019we, myers2018censored}.
\edit{Some journalists have reported on their challenging working conditions as} working up to nine hours a day in what they described as crowded workspaces~\cite{Dwoskin_2019, Soderberg-Rivkin_2023}.

Given the limited \edit{empirical data} on the impact of reviewing \edit{harmful} content on content workers \edit{in the commercialized or professional setting, prior research has turned} to analogous fields to anticipate challenges \edit{RAI} content workers face. 
Steiger et al.~\cite{steiger_psychological_2021} noted similarities between the psychological impact experienced by content moderators and professionals like journalists, emergency dispatchers, and sex-trafficking detectives who witness traumatic scenes, potentially leading to \edit{PTSD}, peritraumatic distress, secondary traumatic stress, and burnout~\cite{feinstein2014witnessing, troxell2008indirect, brady2017crimes, steiger_psychological_2021}. 
Expanding on this analysis \edit{and recognizing the cognitive dissonance that RAI content workers my face of holding adversarial or harmful values along with their personal and societal values, we look to} numerous other occupations \edit{that} confront \edit{similar} moral dilemmas and psychological distress. 
For instance, military soldiers and healthcare professionals often grapple with moral injury when making decisions that conflict with their personal beliefs~\cite{litz2009moral, dean2019reframing, greenberg2020managing, maunder2006long}. 
Similarly, military interrogators, actors, and ethical hackers may face moral conflicts when required to adopt roles contrary to their values ~\cite{konijn2000acting, jordan2017sociology, arrigo2004utilitarian}. 
Social workers navigate bureaucracy and systemic issues, paralleling content workers who must confront distressing content they cannot prevent ~\cite{haight2016scoping}. 
\finaledit{\edit{Our work aims to highlight lived experiences and well-being challenges of RAI content workers in professional and organizational settings and to explore} the unique moral challenges encountered by content workers}.

\subsection{Workplace Support for Well-being and \edit{Harmful Content} Exposure}
In line with the plethora of challenges identified for content workers, emerging literature has examined how content work may be supported from multiple angles. 
Prior studies have explored the benefits and implementations of reducing exposure by limiting the visual field ~\cite{lin2009capture}, limiting color display ~\cite{spence2006color}, limiting amygdala activation from viewing facial emotions ~\cite{costafreda2008predictors, fusar2009functional}, reducing intrusions~\cite{holmes_key_2010}, displaying content in monochrome greyscale, blocking faces, blurring visuals, muting audio, controlling the speed of videos, and allowing content workers to play Tetris immediately after viewing content ~\cite{holmes_can_2009, das_fast_2020, karunakaran2019testing}.
\edit{Other well-being approaches include monitoring} the short- and long-term impact of viewing content ~\cite{watson1988development}, programs to improve resilience skills ~\cite{steiger2022effects}, \edit{evidence-based psychotherapy~\cite{cusack2016psychological,watts2013meta,cook2022awe, spence2023content}}, \edit{and a suite of workplace, clinical, and technological interventions} ~\cite{steiger_psychological_2021}.

While much work on content workers we outlined has focused on content exposure and mitigation strategies, our \edit{work} examines the broader \edit{workplace} context surrounding a content worker \edit{because of the intricate relationship between work, individuals, and their well-being~\cite{robertson2011well, sousa2000well,knight2006impact,biggio2013well,Chari2018ExpandingTP}: work can simultaneously be a source of well-being support via Employee Assistance Programs (EAPs)~\cite{masi2020history}, a source of meaning~\cite{wrzesniewski1997jobs}, and a source of well-being challenges~\cite{stress_2021,colligan2006workplace,grzywacz2002work}.}
\edit{Therefore, in addition to the utilization of exposure reduction tools in their day-to-day work, we examine content workers'} physical work setup, organizational support, and its utilization, as well as their perception toward work and place in the broader context of society. 
We also investigate how \edit{workers} access or incorporate treatment resources, such as therapy, to cope with symptoms of exposure with organizational support.

Automated content regulation, which reduces moderator exposure to harmful content, has \edit{also} been explored through various algorithmic approaches in images, videos ~\cite{Wang2012, Deniz2014, Ries2014}, and text ~\cite{MacAvaney2019, Halevy2020, Schmidt2017, Jurgens2019, park2016supporting, saude2014strategy, hammer2016automatic, gollatz2018turn}. 
These methods often target specific categories of harm, including pornography, pro-eating disorder content, mental health content, personal attacks, and hate speech ~\cite{singh2016behavioral, chancellor2017multimodal, Zhang2023, wulczyn2017ex, hammer2016automatic, MacAvaney2019}. 

\edit{Recently, tools have been developed to increase automation in RAI red teaming~\cite{ mazeika2024harmbench, PyRIT}, raising concerns about their robustness and coverage of discovering harms and effectiveness in reducing harm exposure.}
\edit{In general}, the introduction of automated content work can lead to concerns regarding potential automation errors ~\cite{jhaver2019human} \edit{or exacerbating the problem they aim to solve~\cite{Gorwa2020AlgorithmicCM}}, even acknowledged by platform leadership grappling with nuances ~\cite{Canales_2021}. 
\edit{In our work}, we explore the \edit{benefits and challenges of using automated tools from workers' perspectives and the organizational support for such tools}.

%% file: sections/3_phase1_methods.tex
\section{Phase 1: Understanding the Nature \edit{and Challenges }of Content Work}
To contextualize content work and the well-being of content workers within the space of generative AI deployment and safety, we examined content moderation and data labeling, alongside the emerging practices of responsible AI (RAI) red teaming. 
Our research questions are as follows:
\begin{enumerate}[start=1,label={\bfseries RQ\arabic*}, leftmargin=1.5cm]
    \item \label{methods:rq-nature} What is the nature of content work\edit{, in terms of professional roles, the types of content handled, the work environment, its impact on well-being, and collaborative practices, as experienced by} \finaledit{content workers}? 
    \item \label{methods:rq-challenge} What are the main challenges faced by content workers in their well-being? 
    \item \label{methods:rq-coping} How do we best support the well-being of content workers?
\end{enumerate}

We broadly define ``content work'' as any work activity related to anticipating, generating, reviewing, reasoning about, or making decisions on digital content. To examine the diversity of content-related work experiences, we divided content work into three \edit{role} categories of content-related activities, which we used throughout our study: 
    (1) \textit{Content Moderation} involves reviewing various forms of online content (e.g., text, photos, audio, and video) with the intent to flag or identify any content that potentially violates the platform's policy or guidelines. 
    (2) \textit{Data Labeling} involves reviewing, labeling, or categorizing various types of content (e.g., text, photos, audio, and video) according to specific labeling or sorting guidelines, which aids in data analysis and training machine learning models. 
    (3) \textit{Red Teaming} involves critical assessments of product or platform features by simulating the actions of potential bad actors or testing system vulnerabilities to identify if these features can inadvertently generate or promote content that violates policy or guidelines, thus enhancing the product's security and safety measures. 
In our study, we specifically refer to RAI red teaming, which evaluates AI system outputs for potential harm.

We conducted a two-phase study~(Figure~\ref{fig:studyflow}), focusing on content workers engaging in the three categories of activities above, who were employed in some capacity (i.e., full-time, part-time, contractor) to conduct content-related activities. 
The first phase was formative and aimed at answering our research questions via a survey and interviews with various content workers. 
In this section, we present our survey and interview protocols, our findings from both studies and a set of preliminary recommendations aimed at supporting the well-being of content workers.
The second phase involved validating our findings and recommendations with content workers in a series of workshops. 
In ~\autoref{sec:evaluation}, we present our workshop protocol and a set of refined recommendations. 
The distribution of self-reported content activities and roles across our phases are found in Table~\ref{tab:demographics}.

\begin{table}[t]
  \centering
    \footnotesize
\begin{minipage}[t]{0.25\columnwidth}
    \begin{tabular}{ll}
    \multicolumn{2}{l}{\textbf{Phase 1: Survey}} \\
    \toprule
    \textit{Activity}  & \textit{Count} \\
    \midrule
    CM only            & 27             \\
    DL Only            & 11             \\
    RT only            & 4              \\
    CM and DL          & 20             \\
    DL and RT          & 1              \\
    All                & 4              \\
    \bottomrule
    \end{tabular}
\end{minipage}
\hfill
\begin{minipage}[t]{0.25\columnwidth}
    \begin{tabular}{ll}
    \multicolumn{2}{l}{\textbf{Phase 1: Interview}} \\
    \toprule
    \textit{Activity}  & \textit{Count} \\
    \midrule
CM only             & 5                \\
DL Only             & 2                \\
RT only             & 1                \\
CM and DL           & 8                \\
DL and RT           & 3                \\
All                 & 3                \\
    \bottomrule
    \end{tabular}
\end{minipage}
\hfill
\begin{minipage}[t]{0.3\columnwidth}
    \begin{tabular}{ll}
    \multicolumn{2}{l}{\textbf{Phase 2: Workshop}} \\
    \toprule
    \textit{Self-identified role}  & \textit{Count} \\
    \midrule
Content moderator/analyst         & 7                 \\
Content designer                  & 3                 \\
Data annotator                    & 3                 \\
Operations or enforcement manager & 2                 \\
Volunteer AI tester               & 1                 \\
& \\
    \bottomrule
    \end{tabular}
\end{minipage}
\vspace{1em}
\caption{Number of participants in various self-reported content activities or roles across three studies (survey, interview, workshop) within two phases.}
\label{tab:demographics}
\end{table}

W\subsection{Methods}
Our mixed-methods formative study aimed to answer \ref{methods:rq-nature} and \ref{methods:rq-challenge} by conducting a survey and semi-structured interviews in parallel. 
We used the survey to get a broad understanding of the diversity of experiences and the interviews to obtain a detailed view of how content work was carried out and the challenges associated with it. 
\finaledit{Our study was conducted between June and August of 2023.}

\subsection{Survey Study}\label{methods:phase1:survey}
Survey participants were recruited \edit{through} a sample of \edit{manually validated} contacts \edit{and} referrals via a snowball sampling method. \edit{We reached out to several technology companies that either conducted content work in-house or hired vendor companies, as well as vendor companies who conducted content work for these technology companies. These companies helped expand our reach by sharing the study information with other organizations and appropriate individuals involved in} the content-related activities defined above. This 15-20 minute survey was anonymous, and participation was voluntary\edit{, with} a strong reminder to avoid accidental \edit{employer} disclosure to protect \edit{participant} anonymity \edit{(see Appendix A.1). To encourage participation, we did not collect employer data, acknowledging this introduces some opacity as a limitation.} Participants received a \$10 gift card \edit{via} an independent survey that only captured their contact information for compensation and \edit{interview} follow-up. The lead institution's \finaledit{Institutional Review Board} (IRB) approved our survey study protocol. 

We received a total of 67 complete responses (see Table~\ref{tab:demographics} for activity distribution). 
The median age of our survey population was 30. 
All reported the highest level of education to be at least some degree post-high school diploma (i.e., vocational training, Bachelor's degree, postgraduate degree, and beyond). 
47.8\% of our population identified as female, 49.2\% as male, and 3.0\% as non-binary. 
77.6\% reported as being full-time employees, 20.9\% as full-time contractors, and one as a part-time employee. 
Most content workers had at least two years of experience in their roles, with 22.4\% having more than five years, 31.3\% having 2-5 years, and the rest having less than two years. 
In this paper, we note our survey participants with the prefix S. 

Our survey questions included several sections aimed at understanding the nature and impact of content work: 
(1) \textit{Background:} We obtained basic demographic information (e.g., age, gender, education), employment status, involvement in different content activities described above, and tenure in these activities. 
We asked participants to describe their motivations for conducting current work. 
(2) \textit{Work Description:} We asked participants to report the modality (e.g., text, image, audio, video) and categories (e.g., child abuse, graphic violence, sexual content) of content they worked with, the average weekly hours, and average daily contiguous hours of content exposure.
(3) \textit{Work Tools and Strategies:} To understand tools and strategies used for work, we enumerated 17 work tools and 23 coping strategies based on existing resources outlined in \autoref{sec:background-challenges} potentially used during or after reviewing or generating content and asked whether they had access to them, whether they used them, and how useful they were. 
All tools and coping strategies listed in the survey can be found in Figures \ref{fig:categories} and ~\ref{fig:tools}.
(4) \textit{Work Impact:} To understand how content work impacts well-being, we asked participants to describe the positive and negative impact of work, challenges related to content work, opportunities for supporting their work, and the impact of the recent rise in the use of generative AI technologies. 
We asked participants to subjectively assess the quality of sleep (5-points) and the frequency of nightmares, flashbacks, or intrusive thoughts (7-points)\edit{, as they are relevant symptoms often associated with PTSD~\cite{american2013diagnostic}.
We provide our full survey questionnaire in Appendix A.2.}

\subsection{Interview Study}\label{methods:phase1:interview}

Interview participants were recruited from those who expressed interest in follow-up interviews in the survey or through referrals. 
These participants were then consented to participate in a 1-hour interview study.
Each interview was recorded and transcribed over video-conferencing software. 
We asked participants to report which of the three content-related activities they conducted in their role to ensure diversity in sampling.
Participants were compensated with a \$50 gift card. Our interview study protocol was approved by the lead institution's IRB.

We conducted 22 semi-structured interviews where 16 reported engaging in content moderation, 16 for data labeling, 7 for red teaming, and 8 who identified with just one activity.
The median age of our interview population was 30. 
12 participants identified as female, 9 as male, and 1 as gender nonbinary.
18 were full-time employees, and 4 were full-time contractors.
In this paper, we note our interview participants with the prefix P. 

\edit{Following the aspects of the nature of content work we are interested in (\ref{methods:rq-nature}), we structured interviews into four main sections: }
(1) \textit{Work Setup:} 
\edit{We asked participants to describe their roles, work activities, and the types of content they encountered, followed by their work location, physical setup (e.g., equipment), environment (e.g., noise, comfort), social setup (e.g., access to colleagues, collaboration), and the ideal work setup for them.}
(2) \textit{Conducting Work:} 
\edit{We asked participants to describe how they conduct their content-related work activities, by sharing how they interact with their tools. We explored how automation or generative AI technologies is or could be helpful in their work as a tool used for work (e.g., filtering, labeling, prompt generation) and as materials for work (e.g., AI-generated images).}
(3) \textit{Coping:} We asked participants to describe how their well-being was impacted by their content work\edit{, including their} current and ideal strategies, resources, or support they leverage to improve their well-being.
(4) \textit{Collaboration:} Finally, we asked participants to describe how they collaborated with others for work, including getting help for content-related work or for social support.
\edit{We provide a full list of questions in Appendix A.3.}

\subsection{Survey and Interview Data Analysis}
For qualitative data from open-ended survey responses and interview transcripts, we applied reflexive thematic analysis~\cite{merriam2002introduction} and open-coded ~\cite{charmaz2006constructing} qualitative texts. 
The survey responses were affinity mapped using FigJam\footnote{https://figma.com} by two researchers. 
The interview transcripts were uploaded to Marvin\footnote{https://heymarvin.com/} and open-coded using the tool by the first author on a granular, line-by-line basis and then with higher-level themes in mind in the following coding iterations. 
At the end of the coding process, three researchers iteratively reviewed codes, resolved disagreements, and refined or grouped codes to identify overarching themes. 

\finaledit{For quantitative data from survey responses, we computed percentages of participants that reported different roles, amount of work, exposure to various content types and categories as well as access to tools and coping strategies. 
We conducted} Pearson's correlation \finaledit{where appropriate}.
All correlation results reported in this paper are statistically significant with p$<$0.05.

\subsection{Privacy and Positionality Statement}
Our study focused on the well-being of individuals who self-identify as employed in content work roles. 
As such, we placed great emphasis on acknowledging and addressing potential participant concerns about the privacy and sensitivity of the data with respect to their employment. 
To conduct our investigation, we gathered perspectives from participants across various workplaces. 
To ensure that participants were comfortable during the study, we emphasized that participation was voluntary at any point. 
Additionally, participants were told their responses would remain anonymous, and they were given the option to remove any data they provided to us afterward. 
We especially encouraged participants not to disclose their place of employment or personally identifiable information throughout our survey, interviews, and workshops. 
After collecting all of our data, we carefully de-identified all accidental disclosures of information that could result in identifying a participant. 
Therefore, we did not disclose specific organizations or roles of individuals and instead reported demographic data in the aggregate. 

We recognize that the perspectives and biases we hold as a research team play a role in our study.
Our research team brings interdisciplinary research expertise in HCI, CSCW, critical computing, affective computing, and psychology and comprises individuals with diverse gender, racial, and cultural backgrounds, including people of color and immigrants. 
We acknowledge that, as researchers, we may have influenced the system under study by observing and probing it. 
We provided participants with a space in which to reflect on aspects of their work experiences and ways to improve them, and such an opportunity to voice their perspectives may not have been readily available. 
It is in this way that our research and biases may, in turn, have provided opportunities for understanding and empathizing with participants. 
On the other hand, we note that we bring biases in our interpretations of what content work is because, as researchers, we cannot assume that we understand it as well as those who conduct such work on a daily basis. 

%% file: sections/4_phase1_findings.tex
\subsection{Survey and Interview Findings}
In the following section, we detail our findings addressing our research questions focusing on the nature~(\ref{methods:rq-nature}) and challenges~(\ref{methods:rq-challenge}) of content work.

\subsubsection{RQ1: What is the nature of content work?}
\edit{In exploring the nature of content work, we aim to provide a detailed understanding of the main factors that constitute this field. Content work involves a variety of roles, types of content, \finaledit{impacts, protective tools, }and collaboration practices. 
By examining these elements, we clarify what content work entails and the various factors that influence it. 
This context serves as a foundation for understanding the complexities and scope of content work.}

\paragraph{How content workers define their profession}
Our participants reported performing various content-related activities, including labeling text for harm categorizations, triaging user feedback, deciding on content removal or reporting it as illegal, and generating prompts to assess harmful content and model restrictions.
\finaledit{Participants performing a variety of these activities in both our survey population and interviews described doing so specifically for AI systems or anticipating their work to shift to focus more on AI-related harms. }
Our study found that 37.3\% \finaledit{(25/67)} of survey participants reported performing activities spanning multiple role categories -- content moderation, data labeling, and red teaming -- indicating that the boundaries between these roles are not always distinct.
For instance, 39.2\% \finaledit{(20/51)} of content moderators also did data labeling, and 4 out of 9 red teamers participated in content moderation and data labeling, pointing to a need for a broader categorization of content work as a diverse set of roles. 
\begin{figure}[t]
    \centering
    \includegraphics[width=\textwidth]{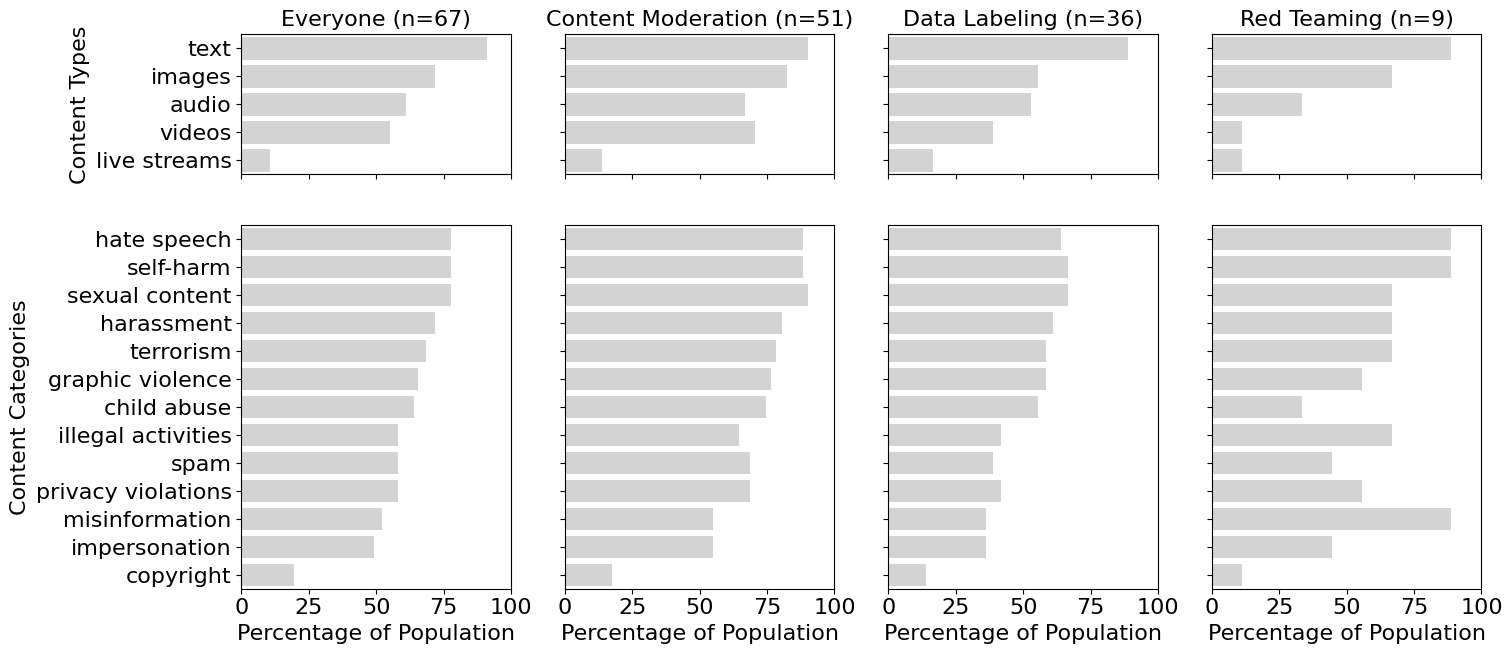}
    \caption{Percentages of our survey population exposed to various content types (e.g., text, images, audio) and categories (e.g., hate speech, self-harm) across everyone (N=67), content moderation (N=51), data labeling (N=36), and red teaming (N=9).}
    \label{fig:categories}
\end{figure}

In addition to the diversity and blurring of roles within content work, we found that content workers \edit{collectively differentiated their roles through professional requirements.
These include self-awareness of personal comfort boundaries with content (e.g., P5, P11), resilience to cope with psychological demands (e.g., P4, P22), subject matter expertise in harmful content (e.g., P6), analytical and investigative skills to interpret human language nuances (e.g., P10, P13), and empathy.} 
For instance, P11 emphasized the need for self-awareness to recruits: \textit{``I want you to be honest with yourself about how willing you are to talk about sexual content, about profanity about religion, about political beliefs, and not only that but to understand the opposing views of those subject matters.''}
Participants highlighted the need to adopt the perspective of harmful content creators (e.g., a white supremacist) to identify \textit{``coded language''} (P10) used to perpetuate bias or to understand how \textit{``people operate differently in those two environments online versus at home''} (P16). 
Red teamers described their investigative processes as intuitive, adaptive, and  developed over time, with P21 emphasizing the need for \textit{``a team of people who are looking forward and anticipating and adapting to the changing circumstances.''}
Overall, our participants highlighted the diverse activities and unique skills required for content work, indicating that this profession may not be suited for everyone.

\paragraph{Types and modalities of content that workers engage with}
We found significant overlap in the types and modalities of content our participants reasoned about, with most reviewing multiple modalities and categories of impactful content.
34.3\% \finaledit{(23/67)} of survey participants reviewed all four modalities (text, images, videos, and audio), while 20.9\% \finaledit{(14/67)} reviewed two modalities. 
Text and images were the most commonly reviewed, with 90.2\% \finaledit{(46/51)} of content moderators, 88.9\% \finaledit{(32/36)} of data labelers, and \finaledit{8 out of 9} red teamers engaging with text, and similarly high percentages for images \finaledit{(82.4\%, 55.6\%, and 6 out of 9, respectively)}.
\finaledit{The content source also varied, as several interview participants (9 red teamers, P18, and P19) reported working with AI-generated content or content within AI systems. }
Moreover, interview participants reported exposure to highly impactful content, such as abusive or hate speech, child sexual abuse material, and terrorist and violent extremist content.
Those involved in red teaming described their work as involving both \textit{``generating''} and \textit{``processing''} such content.   
For instance, P6 described their role as dealing with \textit{``explicit content that is either sexual content or child endangerment content. Violence, racism, the bevy of the worst of the worst of the Internet is what I have to generate and also test, moderate, and sift through.''}
Figure~\ref{fig:categories} shows that half of our population was exposed to all categories of content except for copyright (19.4\%\finaledit{; 13/67}) and others (15.0\%\finaledit{; 10/67}). 
Red teaming participants reported higher exposure to hate speech, self-harm, and misinformation (\finaledit{all 8 out of 9}) compared to the general survey population (77.6\%, 77.6\%, and 52.2\%, respectively). 
Content moderators encountered sexual content more frequently (90.2\%\finaledit{; 46/51}) than the population (77.6\%\finaledit{; 52/67}).
These findings suggest that content workers are frequently exposed to highly impactful content \edit{ regardless of their specific roles.} 

\paragraph{Impacts of performing content work}\label{RQ1:impacts}
Prior research has shown the significant negative impact of exposure to harmful content (Section~\ref{sec:background-challenges}). 
Our findings confirm this but show that symptoms vary by individuals and specific activities.
\edit{We found that the negative psychological impacts persisted long after exposure, resulting in residual effects such as moral injury, lower sleep quality, intrusive thoughts, and hypervigilance.
We found that moral injury~\cite{shay2014moral} stemmed not only from viewing content conflicting with one's moral values or beliefs but also from the adversarial red teaming.}
For example, P6 stated that they had to \textit{``dive into white supremacist blogs...absorb the information of how people talk there, how people utilize language, and then bring that back to the team and process that together.''}
S11 described their work as \textit{``scarring [their] brain for money''} since content work activities \textit{``can be stressful and emotionally impactful,''} highlighting the emotional toll.
We found that the better a content worker red teamed a model, the worse they could feel about themselves for making that model generate harmful output. 

Sleep quality was on average ``fair'' to ``good'' ($\bar{x}$=3.46 out of 5, $\sigma$=1.03), but lower sleep quality was significantly correlated with weekly exposure (Pearson $r$=0.357) and contiguous hours worked per day (Pearson $r$=0.430).
Nightmares or intrusive thoughts were rare ($\bar{x}$=2.56 out of 7, $\sigma$=1.61) and negatively correlated with sleep quality (Pearson $r$=-0.274).
S9 and S19 frequently revisited past choices, experiencing intrusive thoughts and guilt about their work outside work hours. 
Exposure to harmful content increases sensitivity and hypervigilance in daily life.
For example, S47 mentioned being \textit{``unable to watch certain TV shows, movies or even listen to stories or podcasts that involve child endangerment, child crimes, gore, severe violence, mutilation.''} S2 reported feeling less empathy and compassion, while P12 felt both hardened and hypersensitive.

\edit{Participants experienced exhaustion from viewing high quantities of content for extended periods.}
Specifically, 61.2\% \finaledit{(41/67)} reviewed or generated content for 30-40 hours a week, with 44.8\% \finaledit{(30/67)} spending more than four contiguous hours per day (median = 3.5 hours per day).
\edit{Red teamers, on the other hand, were exposed to content for a median of 15 hours per week, with a median of 1.5 contiguous hours per day.}
\edit{Several interviewees, especially those in content moderation, lamented the physical and psychological distress from long exposure to harmful content.} 
For example, P13 mentioned that, \textit{``Sometimes it's nearly a full day's work just going through toxic reports ...it's incredibly draining.''}
\edit{Such findings indicate that the long work hours compounded the other impacts of content work described earlier.}

\paragraph{Tools content workers use for protection}\label{RQ1:tools}
\begin{figure}[t]
    \centering
    \includegraphics[width=\textwidth]{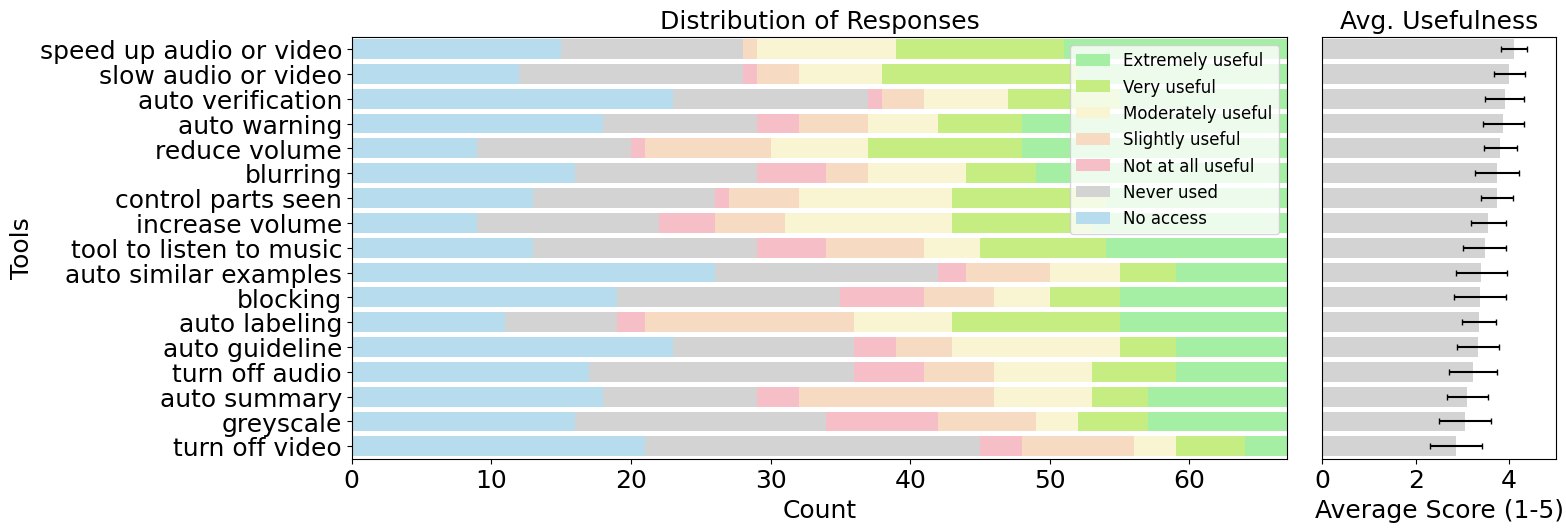}
    \caption{The access and usefulness of tools that help during content work which are sorted in descending order of usefulness \edit{with 95\% confidence intervals}.}
    \label{fig:tools}
\end{figure}

\edit{In terms of} the tools participants used for protection while performing content work activities\edit{, we found that t}he most useful tools were video and audio speed controls (4.1), automatic verification (3.9), and automatic warning (3.9). 
We observed a significant positive correlation between a content worker's contiguous exposure to content and the usefulness of blurring (Pearson $r$=0.340) and blocking faces (Pearson $r$=0.476).
As illustrated in Figure~\ref{fig:tools}, these tools received high average usefulness scores.
 
\edit{Interview participants who red-teamed disclosed access to a differing subset of tools specific to their role, such as those for prompting models. }
For instance, red teamers used a prepared list of prompts that workers would then send as input to models and wait to evaluate model outputs (P19, P20). 
P20 found this tool useful, saying \textit{``the advantage of using this data set would be [reassuring of] `okay, I didn't come up with this.' I just use whatever inputs are there are run them.''}
\edit{Thus, we find that tools used by red teamers may aim to alleviate the internal dissonance they face when generating harmful prompts that go against their beliefs.}

\paragraph{How content workers collaborate as a team}
\edit{The final aspect of content work we explored was its collaborative nature.
Participants benefited from diverse teams, which provided a diversity of perspectives and allowed them to tailor their work experiences to their personal preferences.}
P5 stated that each team member had unique domain expertise, such as vendor management, user interface design, and video editing to \textit{``spot fakes''}, recognizing generated harmful content.
\edit{This diversity enabled P20 to divide responsibilities according to team strengths when red teaming, while P6 typically processed content collaboratively with their team.}
However, P20 cautioned against relying solely on domain expertise, emphasizing the importance of lived experience: \textit{``you don't want a team full of white guys testing the feature. You want everybody there...But that's not the only way, especially if...seeing that repeated offensive information about it can have a long-term effect on you.''}
\edit{Ultimately, some content work activities benefit from collaboration, but the collaborative nature also} \finaledit{opened up challenges in questions about how diversity should be incorporated in a team without overburdening individuals who may have more underrepresented skills or experiences with needing to do more work.}

\subsubsection{RQ2: What are the main challenges faced by content workers in their well-being?}
\edit{In this section, we illustrate key challenge our participants experienced as content workers, which informed our recommendations listed in Table \ref{tab:revised_table}. 
Each subsection details the identified challenges and the corresponding recommendation categories we developed.}

\paragraph{Misconceptions and realities of content work}
Several participants found their initial perceptions of content work to be inaccurate after starting their roles. 
While some viewed it as \textit{``just a job''} (P4, P7, P17), this view was often challenged by the realities of the work's impact. 
P2 reflected: \textit{``I was like, oh, this is going to be a really easy job...And it wasn't easy.''} 
Participants used various methods to cope with the ``not easy'' parts of content work, such as putting up figurative shields (P9), \edit{but these often backfired. }
As P13 noted,  becoming \textit{``hardened to [harmful content] in the sense that it doesn't impact [them] as much...can cloud [their] judgment''} due to frequent exposure.
Many participants expressed the need for adequate warnings about the nature of content work, including the activities performed (P2, P17), types of content (P2, P12, P17), and potential impacts (P17). 
\edit{P19 recommended surveying new workers about their comfort level and offering alternative assignments. 
Training resources, such as video curricula with coping strategies, were also found helpful (P1).}

\edit{From these challenges, we developed our first category of recommendations: \textit{Recruitment.} 
We urge that potential RAI content workers receive comprehensive education about the impact of content exposure. 
This education helps them assess their suitability for content work, which demands a specialized skillset honed through experience. 
Early, personalized training should equip new hires to cope with the highly individual naure of exposure-related symptoms, potentially utilizing tools like a generative training dataset for controlled exposure.
}

\paragraph{Shortcomings of existing tools and metrics}
\edit{We identified several challenges with the tools available to content workers.} 
Many tools were inaccessible to some participants, including automatic clustering (38.8\%\finaledit{; 26/67}), automatic verification or guidelines (34.3\%\finaledit{; 23/67}), turning off video (31.3\%\finaledit{; 17/67}), and blocking (28.4\%\finaledit{; 19/67}). 
35.8\% \finaledit{(24/67)} and 28.4\% \finaledit{(19/67)} of participants who had access to turning off video or audio never used these features.
This is concerning as these tools were particularly useful with greater contiguous exposure to impactful content (see section \ref{RQ1:tools}). 
Participants who regularly used tools recognized the need to address variability in tooling needs. 
For instance, while P13 viewed images unfiltered to understand the context, P7 needed the black-and-white filtering feature.
\edit{Generative AI was mentioned as an emerging tool, but \finaledit{has many areas for improvement in application.} 
For instance, P6 emphasized the need for human involvement due to the focus of content work being on \textit{``the bleeding edge of technology''}, making it difficult to automate red teaming fully.}
\finaledit{As such, challenges remain in reconciling the benefits of generative AI with the drawbacks of it potentially exacerbating the burden on workers by creating more content for workers to engage with. 
P11 anticipates this being a key challenge: \textit{``I have a list of 100 terms, and I use a prompt against it to identify which of these terms is sexual content [...] I run that prompt four times. If each instance of that prompt results in a different result, how do we reconcile that?''}}

Participants also highlighted challenges with how productivity was measured, sometimes facilitated by the tools they use for work. 
Many felt pressured to work long hours because their productivity was measured by time worked or content volume reviewed. 
Instead, they preferred metrics that helped them manage when to take breaks (P1) and recognize the impact of their work. 
P7 noted that being told to view content often worked against their productivity on days with insufficient content to review. 
Participants were more receptive to  a combination of metrics, including content severity or the severity of impact as well as the quantity and time spent viewing content.

\edit{Based on these findings, we recommend the category of \textit{Tooling.}}
Tools should be customized to reduce content exposure based on individual preferences. Continuous feedback from content workers should be incorporated into tool improvement and creation to ensure relevance to their needs.
\begin{figure}[t]
    \centering
    \includegraphics[width=\textwidth]{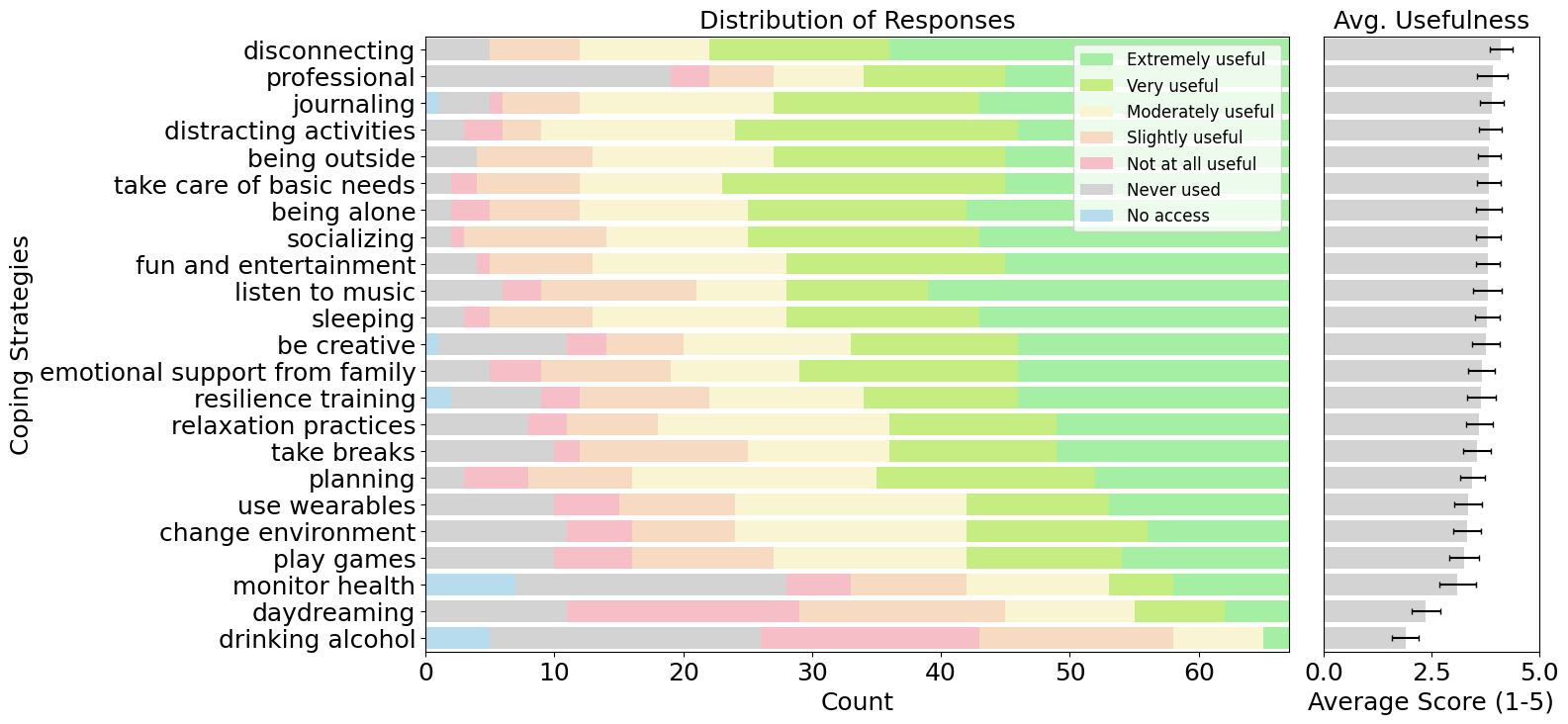}
    \caption{The access and usefulness of coping strategies used to manage the demands of content work which are sorted in descending order of usefulness \edit{with 95\% confidence intervals}.}
    \label{fig:coping}
\end{figure}

\paragraph{Failures in workplace support and coping mechanisms}
\edit{We found that content workers use highly individualized and specialized methods to cope with the psychological demands of their work.}
\edit{Most coping strategies were accessible, but some lacked access to health monitoring (10.4\%\finaledit{; 7/67}) and resilience training (3.0\%\finaledit{; 2/67}).  
Despite all participants reporting having access to professional support services (e.g., therapy), only 71.6\% \finaledit{(48/67)} of participants used them, with only 5 out of 9 red teamers participating. } 
\edit{The most useful coping strategies were disconnecting (\finaledit{$\bar{x}$=}4.1 \finaledit{out of 5, $\sigma$=1.1}), professional support (\finaledit{$\bar{x}$=}3.9 \finaledit{, $\sigma$=1.3}), journaling (\finaledit{$\bar{x}$=}3.9 \finaledit{, $\sigma$=1.1}), distracting activities (\finaledit{$\bar{x}$=}3.9 \finaledit{, $\sigma$=1.1}), and being outside (\finaledit{$\bar{x}$=}3.8 \finaledit{, $\sigma$=1.1}), as shown in Figure~\ref{fig:coping}.}

\edit{Interview data highlighted variabilities in individual preferences.
For example, P7 found work therapy sessions unhelpful and \textit{``[felt] like a waste''} because they were \textit{``not allowed to talk about issues that are bugging [them] from outside of work.''}
In contrast, P3 benefited from workplace coaching. }
P20 found group therapy helpful for shared experiences and found it to be \textit{``a really nice space because [they] get to see what everybody else and going through and [they] realize that sort of a lot of [them are] having the same or similar experiences.''}
\edit{However, we found an under-utilization of professional support services among red teamers, partly due to the nascent nature of their roles.} 
For example, P21, who was satisfied with playing Tetris or taking walks for occasional red teaming and seeing a disturbing image, conjectured that \textit{``if somebody was doing it for six and a half hour days, I would probably say mandatory. Like, I might literally force people to play Tetris if they had to do that much content moderation or something akin to it because, boy, it'd be hard.''} 
P19 did not feel affected enough to join support services, and P18 highlighted that \textit{``[well-being sessions] are not for vendors.''}

\edit{Access to coping strategies was also limited by the work environment.
P9 found home-based strategies, like having pets or listening to music outloud, effective, while others preferred access to colleagues in the workplace for fostering stronger interpersonal relationships.}
\edit{P5 found regular walks with colleagues beneficial for well-being \textit{``check-in''} and venting.
P7 noted challenges of working in the office as well as working from home. They noted being \textit{``blown away in the face''} by the content their colleagues were viewing when they pass by their screens in the office and causing \textit{``second-hand smoke''} when seeking support from their loved ones at home. }
\edit{These findings underscore the need for  \textit{Adaptive Wellness.}}
We recommend flexible and individualized support for content workers, recognizing the varied impacts of their work.
Providing diverse well-being resources and fostering workplace connections are crucial for validating their experiences and offering support through both informal and formal interactions.

\paragraph{Barriers to career growth and support}
\edit{We found that participants faced challenges regarding access to career opportunities and resources, as many lacked an understanding of career growth in content work. 
Most participants viewed content work as an entry-level position with opportunities to advance in fields like user experience design, project management, or software development (P6, P9, S33, S60). }
For example, P16 remained in their role due to advancement opportunities: \textit{``I started showing that skill set to want to dive deeper and to ask more questions...when the interview process came up for the tier two position, I was asked if I was interested in interviewing.''}
Conversely, P5 felt there were no growth opportunities: \textit{``We don't have networking opportunities, and I don't see any growth. The way I grow in this role is I get a new skill set and then leave this role. There's no promotion. There's no career pathway.''} 
\edit{The perception of content work as temporary and replaceable thus contributed to the lack of well-known career opportunities.}

\edit{To remain motivated, some participants sought closure by learning about the positive impacts of their work through reports of the number of children they save (P5) or progress on investigations (P13), but accessing such data was challenging due to the sensitive nature of the content (P2, P4, and P16).}
S39 found it rewarding: \textit{``While it is hard seeing some of the content I am exposed to, it can be a bit rewarding, knowing that you are keeping more vulnerable users safe.''}
\edit{Even then, several participants expressed frustration that the public did not understand the importance of content work.
P5 recounted a discouraging remark: \textit{``Man, your job sounds great. you just look at porn all day.''}
P21 highlighted the lack of positive recognition:\textit{``You only notice [red teaming work] when something goes wrong. You don't notice it when it's going well.''}}

\finaledit{Participants expressed concerns about growing workloads due to the rapid development of AI systems and the insufficient resources for content workers. P20 described the constant pressure on their team, noting, \textit{``given how much we're shipping, how many features we've had to review, and how much [time in] our schedule, we allow for a certain number of [sessions] in a week, and we're always overbooked. ''} Anticipating similar challenges, P18 advocated for hiring more content workers but worries about the lack of resources for current workers. Generative AI and automation have been introduced to reduce workloads and exposure. Some welcomed having automated tools for the purpose of reducing exposure and potentially lessening the negative impacts of their work (P12 and P15). While we anticipated a concern about automation replacing workers, our participants did not share this sentiment. }
\edit{Several participants were confident that advances in automation would not drastically alter their work, believing that such attempts could not replace their roles, saying that \textit{``no amount of automation is going to be able to do the nuanced work that humans can do in this thing''} (P11). 
\finaledit{Crucially, participants knew }that several automated methods do not even apply to activities such as red teaming for generative AI (P13 and P15).
Thus, we confirm that a challenge persists in } \finaledit{supporting content work amid growing demands in AI development while utilizing state-of-the-art technologies such as generative AI to complement existing workflows.}

\edit{Participants also emphasized the need for supportive leadership that understood content work, advocate for and listen to workers (P5, P9).}
P9 argued that stakeholders need first-hand experience: \textit{``it's just one of those things where unless you're actually in it, it is difficult to understand the gravity and impact of the work.''}
\edit{With increasing workloads, the need for strong, capable leaders is urgent.} 
Specifically, red teamers expressed concerns about rising demands (P8, P20), with P8 expressing nervousness: \textit{``We expect this [type of content's] volume to increase significantly, and keeping up with it will be a real challenge. I do not know how to do that, and I'm nervous about that. Everybody else I'm working with is nervous about that.''}

\edit{Overall, we highlight the need for clear career pathways and growth resources facilitated by prepared leadership in the face of increasing workloads. 
Our final category of recommendations,\textit{Retention}, calls for efforts to demonstrate the value of RAI content work and support skill development and retention.} 
Career pathways and transferrable skills learning can facilitate transitions from perceived short-term roles. 
Networking opportunities, such as internal conferences, can promote personal growth and validation.

%% file: sections/6_phase2.tex
\section{Phase 2: Evaluating Recommendations}\label{sec:evaluation}

In the second phase of our study, we aimed to \edit{address \ref{methods:rq-coping}: \textit{``How do we best support the well-being of content workers?''}.
To do this, we validate our recommendation framework and surface persisting challenges and consideration in its implementations through a }series of small-group workshops with those who identify as conducting content moderation, data labeling, or red teaming activities. 
Our workshops are designed to facilitate a form of ``member checking''\edit{~\cite{lincoln1985naturalistic,creswell2012educational,yin2009case} as a way to validate our findings from the first phase~\cite{lincoln1985naturalistic,creswell2000determining} and to provide an avenue for consensus and empowerment, through highlighting participants' voices on the feasibility and implementation of our recommendations within the complex individual, organizational, and societal contexts our participants navigate daily~\cite{cho2006validity}.}
In this section, we present our workshop protocol, our findings from these workshops, and revised recommendations.

\subsection{Workshop Methods and Data Analysis}

The recruitment method for workshops was similar to that of the interviews. 
\edit{Consented participants attended a 1-hour workshop session} conducted remotely over video-conferencing software and FigJam\footnote{https://figma.com}.
We recruited a total of 14 participants across 2 workshops (Table~\ref{tab:demographics}). 
11 participants reported as being full-time employees and 3 as being full-time contractors. 
5 participants identified as female, 8 identified as male, and 1 identified as transgender. 
The median age range of our workshop participants was 36-45 years old. 
All reported the highest level of education to be at least some degree post-high school diploma (i.e., vocational training, Bachelor's degree, postgraduate degree, and beyond).
Participants were compensated with a \$50 gift card. 
Our \edit{workshop} study protocol was approved by the lead institution's Institutional Review Board (IRB).

\edit{We kicked off each workshop with a quick introduction and tutorial on FigJam (10 minutes). In the first portion of the workshop (20 minutes), we asked participants to review challenge themes we discovered in phase 1 and vote on those that they agree or disagree with, comment on whether they think the challenge is accurate, react to each other's notes, and discuss emerging themes. In the second portion of the workshop (30 minutes), we asked participants to review the AURA framework with its preliminary recommendations from phase 2, comment on what they liked, what they didn't like, what they would change, how the recommendations could be improved, how they would want the recommendation implemented, and discuss emerging themes. Detailed structure of the workshop, including the challenges and recommendation cards and screenshots of the interactive boards presented to the participants, can be found in the Appendix A.4.}

\edit{We applied reflexive thematic analysis~\cite{merriam2002introduction} on workshop responses, both written and spoken. }
\edit{Participant reactions were} systematically collated \edit{to assess} the accuracy of challenges by ascertaining the extent of consensus or divergence among participants through voting. 
\edit{We further} analyzed the remaining qualitative data by surfacing themes \edit{in} responses to our proposed recommendations.
In this paper, we note our workshop participants with the prefix W. 

\subsection{Workshop Findings}
\edit{In the following section, we present our participants' discussions on the identified challenges and concerns regarding our recommendations. Overall, our findings validate the AURA framework and its four pillars. 
They emphasize the importance of our recommendations and highlight participants' concerns about their implementation. 
These insights are integrated into our revised table of recommendations (Table ~\ref{tab:revised_table}
) to enhance the comprehensiveness and impact of our guidance.}

\begin{table}
    \centering
    \begin{tabular}{|c|c|p{7cm}|p{7cm}|}
        \hline
        \rotatebox[origin=c]{90}{} & \multicolumn{2}{|c|}{Recommendation} & Example Applications of Recommendation \\
        \hline
        \multirow{2}{*\rule{0pt}{15pt}}{\rotatebox[origin=c]{90}{Recruitment}} & R1 & Ensure potential participants of RAI content work are informed about the psychological impact of content exposure. & Providing descriptions of types of content that are present in the datasets within content work job descriptions \\
        \cline{2-4}
        \rule{0pt}{15pt} & R2 & Educate content workers with basic and ongoing training to minimize exposure and ease psychological burdens associated with content work. & Providing AI-generated examples of each type of content a new hire will be exposed to \\
        \hline
        \multirow{2}{*\rule{0pt}{15pt}}{\rotatebox[origin=c]{90}{Tooling}} & R3 & Allow configuration of tools to accommodate individual sensitivity to content exposure. & Limiting content workers to viewing one case at a time (e.g., one image at a time) \\
        \cline{2-4}
        \rule{0pt}{15pt} & R4 & Perpetually integrate feedback into tools to stay responsive to the evolving demands of content. & Providing a platform for bug and feature requests \\
        \hline
        \multirow{2}{*\rule{0pt}{15pt}}{\rotatebox[origin=c]{90}{Adaptive Wellness}} & R5 & Provide flexible well-being support to accommodate highly variable responses to content exposure. & Providing a craft table for content workers to engage in creative hobbies (e.g., origami) \\ [2ex]
        \cline{2-4}
        \rule{0pt}{15pt}& R6 & Foster strong interpersonal work connections that can act as first-aid to the psychological burdens of content work. & Organizing optional team lunches with food paid for on a monthly basis \\
        \hline
        \multirow{2}{*\rule{0pt}{15pt}}{\rotatebox[origin=c]{90}{Retainment  }} & R7 & Enable content workers to observe the beneficial effects of their work, mitigating psychological stress resulting from feedback absence and demonstrating the value of their contributions. & Sending a weekly newsletter with a digest of a team or individual’s contribution to keeping a platform safe \\
        \cline{2-4}
        \rule{0pt}{15pt} & R8 & Design well-defined career pathways for content workers to foster the retention of domain experts. & Organizing a conference on responsible AI for content workers to network with peers across teams and organizations \\
        \hline
    \end{tabular}
    \caption{Revised AURA framework and its eight recommendations to support RAI content workers, categorized by the framework's four key categories of holistic support.}
    \label{tab:revised_table}
\end{table}

\subsubsection{Agreement with Discovered Challenges}
As we delved into the challenges unearthed during the initial phase of our study, we found participants actively engaged with these challenges and envisioned their personal relevance. 
Participants most strongly agreed that six of the eleven challenges we identified were accurate: psychological and physical symptoms from generating, analyzing, or researching harmful content, lack of career opportunities to grow in content work, an inaccurate measure of productivity or exposure, lack of specialized tools for content work, lack of appreciation for content work, and lack of closure on the positive impact of work.
W3, \edit{reflected on} how accurate the challenge of lack of career opportunities was: \textit{``the skills [they] develop does not feel like it opens many opportunities nor are there many closely adjacent roles. As such, there's a limited number of options to progress, mostly into people management positions''}. 
On the other hand, participants least strongly agreed with the accuracy of three challenges: the double-edged sword of heroism, moral injury from working with harmful content, and inaccurate expectations about the cost of content work. 
\edit{For instance, while W10 felt \textit{``[they were] 100\% aware of what [they were] signing up for,''} they still noted the inaccuracy in job descriptions for new hires and advocated for the implementation of a recommendation to address this. 
As such, even when participants did not experience the challenges themselves, they valued the recommendations for others facing these issues.}

\subsubsection{Implementation of Recommendations}
Through discussions with participants, we surfaced \edit{many} concerns regarding how the recommendations should be implemented in practice. 
We integrated these perspectives into our revised recommendations, providing detailed considerations in the subsequent section \edit{and updated examples illustrating our recommendations' application}.
We also note that participants did not propose additional recommendations beyond those we presented; instead, they expressed concerns with a critical focus on how our recommendations may be implemented.
\edit{When considering hosting a conference for learning and networking, W14 }exclaimed that \textit{``If this could be shared from [their] managers to the team, it would be helpful.''}
\edit{Conversely, W13 expressed concern that attending such events would take away from their assigned production hours--current productivity metric--to attend such an event, making them less likely to go.
From another perspective, W1 suggested making such meetings \textit{``a peer-to-peer meeting style that allows for a collective group of content moderations/designers/researchers to nominate an agenda.''}
These concerns highlight the need for organizational and managerial support because, as our participants reported, content workers often lack the resources to initiate such changes.
Further from W1's insights, we find that organizers should carefully account for workplace dynamics to ensure content workers are in spaces they deem productive for growth. }
\edit{However, there were also }recommendations that content workers could support \edit{independently.}
For instance, W10 stated that \textit{``[their] team is currently interviewing candidates for an open role. The job description is absurdly vague and doesn't accurately describe the work or its impact,''} \edit{highlighting an opportunity for content workers to improve job descriptions.
Similarly, W11 shared how they provided feedback on the tools they used, demonstrating content workers' agency in implementing our recommendations.}

\subsubsection{Considerations for Recommendations}
\edit{In the next section, we synthesize participants' insights into considerations for each pillar of the framework, along with our revised recommendations. We include examples from workshop participants to illustrate each consideration.}

\textit{Recruitment}: We found that careful consideration should be placed to ensure that recruitment \edit{involves more than a }single training instance. 
W3 stated that this recommendation \textit{``should be more than recruitment but ongoing training with new information.''}
In fact, W3's point highlights how new training should be provided continuously so content workers' skills for well-being and their work may be kept up to date with the evolving demands of content work.
We emphasize in our recommendation (R2) that content workers are educated with not only basic training at the initial recruitment stage but provided\textit{ongoing} training throughout their work. 

\textit{Tooling}: 
Participants predicted that implementing tooling guidelines may distribute the responsibility of ensuring the effectiveness of such tools to engineering teams (W13).
As such, engineering teams need the proper support (e.g., allocating a portion of their time to maintaining tools) to integrate the feedback they receive from content workers about performance issues and feature requests.
Per W2:
\begin{quote}
    \textit{``[doing so] could allow for quick resolution of technical issues with content moderation tools, as well as minimize the psychological burden of simultaneously dealing with gruesome content and managing the stress of hitting [their] production hours--which may arise due to issues with tooling.''}
\end{quote}
As per our participants' advice, we advocate for engineering teams to be viewed as crucial for content work and efforts for AI testing. 
We urge decision-makers to \edit{allocate sufficient resources to support these teams and, by extension, the content workers who depend on them.}

Additionally, workshop participants emphasized that AI tools assist content workers within their existing workflows. 
W1 \edit{noted that tools allowing }free-form user input for unique content cases to report \textit{``could limit the work [they] do, pushing it onto their co-workers.''}
\edit{As such, we advise that tool improvements involve direct collaboration with content workers, as they are the primary end-users.}

\textit{Adaptive wellness}: 
Concerns appeared about how flexibility in well-being support may unfairly distribute team responsibility. 
\edit{W2 worried that it }\textit{``leaves it up to [the content workers] to create the space rather than it being provided as part of the job.''}
W13 \edit{added} that \textit{``[we] need guidelines on what is acceptable among teams so as to minimize opportunities for bias among employees''} while W3 pointed out \edit{the need to specify management responsibilities}. 
We advise that applications of our recommendations for adaptive wellness are implemented with clarifications on where the responsibilities lie for each team member.
For instance, \edit{it should be clear if content workers must provide materials for a craft table, as in the R5 example.}

Additionally, W3 expressed concern about the cost of more flexible well-being as a limitation. 
\edit{To address this, we recommend that such support be part of workplace provisions and not replace external benefits. 
Participants also worried about activities becoming ``mundane'' and desired alone time (W2). 
Thus, fostering strong interpersonal work connections through optional team events may be beneficial when offered in variation.}

\textit{Retention}:
\edit{
Several workshop participants expressed concerns about lacking time and support for skill development within the workday (W13, W14). 
For instance, W13 worried: \textit{`` This suggestion would take away time from my assigned production hours, in which case I would be less likely to attend [career development events] even if I felt like it was beneficial to my personal growth.''}
Thus, we revised our recommendations to suggest that content workers have allocated time to focus on career growth on at least a monthly basis. }
Ultimately, our participants' immediate focus on applying our recommendations surfaces a sense of urgency they feel as professionals in the RAI content work space to receive such support.

%% file: sections/7_overall_discussion.tex
\section{Discussion}
In this paper, we explored content work involved in the development of RAI technologies. \edit{We identified recent challenges faced by content workers to inform the future design and support for content work.} 
We confirmed that the exposure to harmful content that is demanded in these professions has a profound \edit{and long-lasting} psychological impact \edit{on workers, as described by} prior studies of content moderation \edit{on social media platforms}~\cite{roberts2016commercial, pinchevski2023social, ruckenstein_re-humanizing_2020}.
\edit{In addition, our work revealed nuanced challenges about the profession, including misconceptions about the work, lack of adequate workplace tools and support, and barriers to career growth.}
Our work encompassed \edit{the examination of} various roles that may be exposed to harmful digital content\edit{, including the nascent RAI red teaming role, where we discovered nuanced differences.}
\edit{
For example, the generative and adversarial aspect of RAI red teaming added another layer to the moral injury and introduced the need for prompting tools. We posit that the nascency of red teaming as a role \finaledit{(i.e., it has not been firmly established as a full-time profession at the time of our study)} could attribute to the fewer number of content hours or the under-utilization of support services in comparison to other content roles.
}
\edit{At the same time, we discovered heightened nervousness around rising demands for red teaming.}
\edit{As the workload of all content workers increases amid growing demand within the tech sector and particularly for that of generative AI capabilities~\cite{The_White_House_2023, Perrigo_2023b,brundage_toward_2020, solaiman_evaluating_2023}, we urge the community to} heed the warnings from prior research on content moderation~\cite{ruckenstein_re-humanizing_2020, steiger2022effects, pinchevski2023social} in light of new and old \edit{forms} for content work.
We must develop a comprehensive strategy for establishing resilient human infrastructure that safeguards and supports these content workers\edit{, and our recommendation framework is just an initial step}.

\subsection{Practical considerations}
Prior research has focused on the psychological impact of content work\edit{, recommending} strategies to prevent, reduce, and treat exposure to harmful content ~\cite{holman_medias_2014, das_fast_2020, karunakaran2019testing, watson1988development, steiger_psychological_2021, steiger2022effects}.
Our findings corroborated these strategies, \edit{leading us to incorporate} \textit{Adaptive wellness} and \textit{Tooling} to address individualized psychological impacts.
In these pillars, our findings revealed that organizations should indeed provide benefits such as psychological support services, but this organizational support should go beyond traditional Employee Assistance Programs (EAPs).
We found that simply having resources available is inadequate; organizations \edit{should} encourage utilization, provide personalize\edit{d support}, and \edit{maintain feedback loops}. 
Additionally, we found that organizational and professional support \edit{are crucial for recruitment} and maintain\edit{ing} healthy career trajector\edit{ies}. 
For these reasons, our recommendations included \textit{Recruitment} and \textit{Retention} as two of the four key pillars, emphasized \edit{by our workshop participants}.

Informed by the lived experiences of content workers, our AURA framework \edit{encompasses these} four pillars, designed to holistically support \edit{content workers' psychological and professional well-being}.
Therefore, we recommend that future work on content worker support or future ideation of recommendations cover all four pillars covered in the AURA framework. 
Here, we discuss practical considerations based on our findings for implementing our recommendations.

\subsubsection{Importance of RAI content worker involvement}
\edit{Our validation workshops helped discover} practical considerations for implementing our recommendations. 
We \edit{identified the need for} learning and networking opportunities \edit{and the importance of integrating these into workers' responsibilities through organizational leadership support.}
The workshops also fostered discussions of ``who'' should participate in implementing the recommendations and ``how'', \edit{with} workers identifying their own roles in promoting awareness and transparency of their work\edit{, such as} writing job descriptions. 
This exercise \edit{helped content workers reflect} on \edit{both} the organization's \edit{and their roles} in improving well-being of their prospective hires, their team, and their organization. 
We posit that there may be an opportunity for worker empowerment through participation where employees can exercise direct and indirect control over their work environments~\cite{Marchington1991NEWDI}. 

Based on this experience, we recommend that researchers and organizational leaders include worker perspectives in the iterative design and discussion of organizational processes and policies, as  demonstrated by prior work \edit{using} codesigning methods for integrating RAI practices~\cite{Madaio2020CoDesigningCT}.
A recent study reported that organizational benefits and practices \edit{should be viewed as investments in employee} well-being\edit{, not just business costs}~\cite{Singer2020EmployersRI} \edit{since} work-related stress \edit{has} a profound impact on the business~\cite{Hassard2018TheCO}.
\edit{This perspective is crucial for} content workers \edit{motivated by} heroism mentality but \edit{lacking} adequate support and \edit{recognition}.
In this regard, \edit{organizations must} consider \edit{employee involvement and participation} in evaluating human resource management practices, as well as in the organizational design and decision-making processes, \edit{to promote} greater acceptance of change~\cite{Ullrich2023EmployeeIA} and work satisfaction~\cite{Guest2002HumanRM}.
Although our study focused on workers\edit{' perspectives}, implementing these recommendations is a design exercise~\cite {Junginger2013DesignAI} that \edit{involves} multiple stakeholders (e.g., workers, managers, organizational leaders, human resources). 
As our findings suggest, individual differences in content work \edit{and well-being needs} introduce complexities of interpreting and supporting well-being~\cite{Kawakami2023SensingWI}, \edit{that} top-down policies cannot \edit{address}. 
Careful stakeholder \edit{involvement} can lead to ideal implementations by ``deliberately eliciting potential tensions that occur when stakeholders' values conflict~\cite{Park2022DesigningFA}.''

\subsubsection{Importance of socio-ecological perspective on RAI content work}
Our study also revealed the importance of \edit{considering} workers' environment in understanding and supporting \edit{them}. 
We found that content workers' physical setup\edit{, such as} having access to musical instruments \edit{or outdoor spaces, and social environments, like team support and managers who advocated for them}, directly influence their ability to cope. 
Beyond work, we found that content workers carefully crafted boundaries around their loved ones and society \edit{to manage work's impact, considering both themselves and others,} based on prior interactions involving the content and \edit{their work} experiences.
We also saw content workers struggle with heightened attention on mitigation failures and less celebration of successes, which impacted their perception toward work and well-being. 

Throughout our study, we found that \edit{content workers'} well-being was never just about the exposure to potentially harmful content; \edit{external factors also play a role}.
In fact, many of the decisions made outside of work have a trickle-down effect on the workers themselves~\cite{PAI_ResponsibleSourcing}.
\edit{In relation,} prior research \edit{shows} that workplace well-being involves interaction between individual characteristics and the surrounding environment~\cite{biggio2013well,Richard1996AssessmentOT,Stokols1992EstablishingAM,Ettner2001WorkersPO}.
Therefore, future research \edit{should adopt a} socio-ecological approach to analyze worker well-being and the effects of program implementations.

\subsubsection{Importance of holistic tooling for RAI content work}
The four pillars in the AURA framework provide a \edit{structure for ideating and categorizing} new and existing technology innovation opportunities, including automated tools, \edit{to support content workers holistically}.
Unlike prior research \edit{focusing} on algorithmic decision support or content filtering~\cite{saude2014strategy, park2016supporting,  MacAvaney2019}, our participants \edit{suggested technologies extending} beyond \edit{daily tasks to include} orientation, work management, and analytics.
\edit{Examples include} using generative AI technologies to simulate content exposure to increase understanding of what the job entails for new or potential hires, \edit{dynamic adjustment of breaks using} affective and ubiquitous computing technologies, \edit{and} tracking the positive impact of their work via a digital dashboard to have a sense of closure in their work.
In designing such tools, however, prior research has urged for incorporating contextual factors surrounding the work \edit{to avoid being} a nuisance~\cite{Yang2019UnremarkableAF}.
Therefore, we urge future technology innovation research to be inclusive of all aspects surrounding the content work, not just the content exposure or the direct impact of tools.

\subsection{Labor considerations}
Our study examined the human labor involved in the review and refinement of potentially harmful content generated by AI and the digital ecosystem.
With a renewed interest necessitated by the proliferation of generative AI technologies, we also examined an emerging form of labor we called ``RAI red teaming'', which rapidly evolved and materialized right before our eyes as we were preparing this paper~\cite{friedlerai,FrontierModelForum}.
\edit{Our work highlighted workers' perspectives on} defining what this work is, who should be doing this work, and how to support that work.
\edit{However, many questions still remain around the blurred roles within content work that includes RAI red teaming}, all embedded within the persistent invisibility of such necessary content work.
Here, we discuss our findings that may guide future research in understanding and supporting RAI content work.

\subsubsection{Defining the work}
In our study, we used three categories of content work -- data labeling, content moderation, \edit{and }red teaming -- to define our research scope. 
We found that \edit{while these} categories fit within RAI practices, individual workers may participate in a multitude of activities\edit{, making}  strict categorization less meaningful.
Such diversity in activities, coupled with individual characteristics, \edit{was observed alongside diverse work experience, work structures, resources availability and utilization, and personalized coping strategies}.
We saw correlations \edit{between} well-being outcomes \edit{and the types and amounts} of exposure.
There \edit{was an overlap between traditional content moderation or data labeling and} RAI red teaming, potentially due to the \edit{expanding scope of work required by} generative AI deployments, as some of our participants eluded to.

\textit{So, what is becoming of RAI content work, and where is it going?}
Such blurring of \edit{content work boundaries} is important to monitor, especially when \edit{it involves workers who are} not prepare them for potentially harmful content exposure. 
We hope \edit{these boundaries will stabilize} over time \edit{to protect worker} well-being \edit{and} urge organizations to conduct longitudinal studies to observe how RAI-related content work\edit{, particularly RAI red teaming,} takes shape within and around RAI practices.
\edit{It is important to monitor the placement of these activities within} the overall AI lifecycle (e.g., design, development, or deployment phases) \edit{and their} long-term effects. 
\edit{Since our} study \edit{uncovered} subjective experiences of moral dissonance, future research should systematically monitor moral injury and negative self-appraisal \edit{among} those actively engag\edit{ing} in generative adversarial activities \edit{compared to} those only \edit{viewing AI-generated} content.

\subsubsection{Invisibility of work}
Our findings confirm the prior observation of the invisibility of content work~\cite{carmi2019hidden, roberts2017content,gray2019ghost,spektor2022ai, steiger2022effects} that reported de-humanizing the work by reducing workers' role into ``a human cleansing device''~\cite{ruckenstein_re-humanizing_2020,carmi2019hidden}.
\edit{
Unfortunately, our participants reported that workers are made invisible through others underestimating their importance (i.e., not noticing red teaming until a model has harmful output) and overlooking what content work actually involves (i.e., thinking content moderation for high-risk content is simply watching porn all day).}
While some suggest that the misperception of content work may be attributable to its relatively unskilled and rote nature~\cite{steiger_psychological_2021}, our findings contrarily highlight that the content work requires highly specialized skills of resilience, analytical thinking, and domain expertise. 
Content workers' ``pride''~\cite{seering2022pride} arises from their noble mission of safeguarding others and their exceptional abilities to manage harmful content \edit{while balancing their own} well-being, and it should not be attempted without the proper preparation or qualifications.
The lack of public awareness, stigmatization of the work, and measures to increase productivity deeply impact the psychological well-being of \edit{content workers}~\cite{steiger_psychological_2021}.
\edit{
This concern extends to red teamers, who face similar challenges due to the nascence of their activity and the lack of visibility into what their work entails~\cite{FrontierModelForum,Cattell_2023, human_intelligence2023}.}
\finaledit{Thurs, future research should examine \textit{different ways in which we make content work further invisible}.} 



In the context of generative AI technologies, we cannot ignore the continuous hype of ``automating away'' content work~\cite{Leike_Sutskever_2023, gray2019ghost}. 
\edit{Our participants welcomed automated tools that can protect them from harmful content and integrate seamlessly into their workflows in a way that preserves control over their tools and their decision-making~\cite{Gorwa2020AlgorithmicCM}. }
\edit{However, while automation efforts are often motivated by the desire to protect workers from harmful content, it is crucial to examine whether these efforts legitimize the value of content work or undermine human skills.}
\edit{Some assume} that using people is a temporary stop-gap solution \edit{until automation can take over entirely}: at first, domain experts \edit{might} label and generate harmful content~\cite{OpenAI_2023}, \edit{then} crowd workers~\cite{ganguli_red_2022,perez_red_2022,thoppilan2022lamda,xu2021bot} and offshore vendors \edit{might take over}~\cite{Perrigo_2023b}, \edit{with the eventual goal of full automation}.
\edit{However,} many scholars \edit{argue that} human involvement will remain essential even with advancements in automation~\cite{steiger2022effects, Gorwa2020AlgorithmicCM,gray2019ghost}.
\edit{Emerging tools for red teaming, such as those generating datasets of red teaming prompts ~\cite{radharapu2023aart}, fine-tuning prompts through various approaches (e.g., search-based) ~\cite{feffer2024red}, or assisting} RAI practices~\cite{Radharapu2023AARTAR,Buccinca2023AHAFA}, must be promoted to transform and elevate content work rather than replace and obscure \edit{human contributions}~\cite{ruckenstein_re-humanizing_2020}.

\subsubsection{Supporting the workforce}
Our current study focused on content workers employed by organizations whose responsibilities included RAI-related content activities. 
The challenge of fostering and growing these highly skilled professionals still remains, as the content profession is sometimes considered a ``dead-end career''~\cite{Newton_2020b}.
For example, none of our workshop participants, \edit{including} subject matter experts, \edit{could} imagine \edit{career progression beyond} managing other content workers.
In anticipation of increased demand for RAI content work, \textit{how should we think about the content work profession and their career growth?}
Some suggest that content work should be time-bound~\cite{Newton_2020}, \edit{serving} as a launching point for \textit{``better job opportunities''} (P16). 
\edit{Many} participants reported that \edit{their} specialized skills are difficult to translate to other professions, \edit{suggesting the need} to expand their skill sets to those transferrable to other \edit{fields}. 
However, viewing the content profession as temporary may reinforce its perception as unskilled and rote~\cite{steiger_psychological_2021}. 
Therefore, future research must focus on developing a variety of career pathways within and beyond content work. 

\edit{Formally employed c}ontent workers \edit{are more likely to receive} EAP benefits, as all of our survey participants reported access to professional support services.
However, the workforce \edit{ensuring} responsible \edit{AI} deployment extends beyond \edit{formal employees to} expert volunteers~\cite{OpenAI_Red_Teaming_Network_2023} \edit{and} crowdsource and gig workers~\cite{hettiachchi2019towards,ganguli_red_2022}, who may \edit{lack} access to support services.
\edit{Public events (e.g.,~\cite{Chowdhury_2023,Cattell_2023,human_intelligence2023}) may offer mental health services, but barriers like stigma and lack of awareness can impede meaningful utilization to available resources~\cite{druss2008trends,Mojtabai2011,eisenberg2009stigma}.}
\edit{These examples show that efforts to expand participation in RAI content work have begun, and it is urgent to understand and provide the support this new population of content workers needs.}

\edit{The push to conduct RAI activities and} ``crowdsource a diverse set of failure modes''~\cite{Parrish2023AdversarialNA} \edit{raises} the questions: \textit{who should be doing this work\edit{, and are we} providing adequate support?}
Considering that our findings highlight well-being challenges despite available tools and resources, we must carefully design \edit{recruitment and inclusion strategies.} 
\edit{This includes considering the} sustained engagement \edit{of and support for workers} without professional support services or short gig\edit{s} without adequate training and preparation. 
Organizers of \edit{content work events involving emerging activities, where labor implications have yet to be fully understood,} should not ignore the impacts of RAI content work and \edit{the need for} upfront \edit{and ongoing support for the} workforce.

\subsection{Limitations}
Our study has several limitations that need acknowledgment. 
We primarily focused on analyzing employed content workers, so readers should be cautious about applying our findings to volunteer and crowdsourced workers, as issues like retention might not be relevant to gig workers with one-off tasks, and well-being benefits may not apply to non-full-time employees. 
As our participants alluded to, there is a general perception of content work being outsourceable, and we have already seen it being crowdsourced ~\cite{ganguli_red_2022,perez_red_2022,thoppilan2022lamda,xu2021bot}. 
Future research should aim to provide holistic support for the entire content workforce. 
Additionally, our study relied on the experiences of 96 participants, a small subset compared to an estimated 100,000 commercial content moderators ~\cite{steiger_psychological_2021}. 
Our sample may be biased toward those who have persevered in content work, so further research should strive to amplify the voices of those who have left the field. 
Our recommendations are validated via workshops but were not explicitly implemented in an organization to evaluate its efficacy. 
Lastly, our study is limited due to the field of RAI red teaming rapidly evolving, even within months of our research and writing. The definition put out by The Frontier Model Forum~\cite{FrontierModelForum} was not available when we first launched our study in June of 2023. We hypothesize many changes from the submission of this paper to the eventual publication that may impact the interpretation of our results.
Regardless of these changes, we recommend integrating research on improving content workers' well-being into AI deployment safety efforts from the outset.

%% file: sections/8_conclusion.tex
\section{Conclusion}
As national leadership and industries increasingly call for safer AI systems, it becomes crucial to consider how the humans behind such efforts are supported.
Through this study, we establish that RAI content work requires expertise that cannot be developed by anyone or instantly. 
Individualized support is necessary for those engaging in RAI content work, and this support should cover aspects of \textit{adaptive wellness, tooling, recruitment, and retention}. 
We consolidate our recommendations for holistic support into a framework for amplifying understanding, resilience, and awareness for RAI content workers (AURA). 
Our recommendations are validated and enhanced with examples of potential applications from a series of workshops. 
Ultimately\edit{, our approach surfaces a critical need to address existing challenges with content work amid increasing demands for it, particularly given the increased interest in AI-related advancements}. 
We urge the reader to consider that offering such support alone is inadequate; support should be actively promoted among every level of organizational structure, from content work teams to leadership. 
In turn, we call for these support structures to be in place as part of RAI deployment safety efforts.